\documentclass[aps,pre, groupedaddress, 12pt]{revtex4-2}
\pdfoutput=1
\usepackage[dvipsnames]{xcolor}
 \usepackage{graphicx}
 \usepackage{amsmath}
\usepackage{hyperref}
\usepackage[utf8]{inputenc}
 \usepackage{mathrsfs}
\usepackage{amssymb}
\usepackage{amsmath}
\begin{document}
\title{Clausius' theorem and the Second law in the process of
isoenergetic thermalization}
\author{Vansh Narang$^{1}$}
\email[e-mail: ]{mp22009@iisermohali.ac.in}
\author{Renuka Rai$^2$}
\email[email: ]{rren2010@gmail.com}
\author{Ramandeep S. Johal$^{1}$ }
 \email[e-mail: ]{rsjohal@iisermohali.ac.in}
 \affiliation{$^{1}$Department of Physical Sciences,
 Indian Institute of Science Education and Research Mohali,
 Sector 81, S.A.S. Nagar, Manauli PO 140306, Punjab, India}
\affiliation{$^2$Department of Applied Sciences, 
University Institute of Engineering and Technology
 (U.I.E.T), Panjab University,
Sector 25, Chandigarh 160014, India} 
\begin{abstract}
Isoenergetic thermalization amongst $n$ bodies
is a well-known
irreversible process, bringing the bodies
to a common temperature $T_F$ and
leading to a rise in the total
entropy of the bodies.  
We express this change in entropy using 
the Clausius formula over a reversible
path connecting $T_F$ with $T_f$ which 
corresponds to the entropy-preserving temperature 
of the initial nonequilibrium state.
Under the assumption
of positive heat capacities of the bodies,
the Second law inequality simply follows from
the fact that $T_F > T_f$. 
We extend this approach to the continuum
case of an unequally heated rod, illustrating 
with the special case of 
the rod with constant heat capacity and 
a linear temperature profile.
An interpolating profile between the discrete and 
the continuum models is studied whereby $T_f$,
given by the geometric mean temperature over
$n$ elements, 
is shown to approach the identric mean
of the lowest and the highest temperatures as 
$n$ becomes large.
We also discuss the case of negative heat
capacity in a two-body set up where
 isoenergetic thermalization  
may be forbidden by the Second law.
However, the alternate scheme in which 
first work is performed reversibly on the system and
then an equivalent amount of energy is extracted
in the form of heat, brings the system to an energy-preserving
common temperature.
\end{abstract}
\maketitle
\par\noindent
\section{Introduction}
Consider the problem of change in the entropy of
a body which is brought from an initial
to a final equilibrium state, via
an irreversible process \cite{Zemanskybook}. Since entropy is
a state function, it suffices to calculate
the difference in the entropy of the initial
and the final states. This is done by applying
Clausius' theorem over a suitable reversible path connecting
the given terminal states .
However, if the body is initially in a
non-equilibrium state that is brought to a final equilibrium
state, then there is no single reversible
path to connect the initial  and
final states of the body.
In order to apply Clausius' theorem to this more general
setting, one may consider
 {\it small} parts of the body (denoted below as elements)---each of which is
initially in a local equilibrium state
that can be connected to the final equilibrium state
via a suitable reversible path.

As a typical example of the non-equilibrium state,
we consider a body whose different elements
 are at different initial temperatures.
The body, as a whole, is energetically isolated from the surroundings.
Further, different elements
do not exchange any work with each other so that the energy
exchange between the elements is in the form of heat only.
As heat flows from warmer to cooler elements,
 their temperature differences diminish and the body
 attains a final equilibrium state with a uniform temperature ($T_F$).
This process of thermalization is intrinsically irreversible,
whereby the final
entropy of the body is greater than its entropy
at the start of the process.

It is thus interesting to investigate whether, starting from a nonequilibrium state,
we can use Clausius' theorem in its original
setting to prove the increase of entropy
at a given energy while achieving a final equilibrium
state at temperature $T_F$.
Such an approach has been recently
highlighted, by one of the authors, for systems having
positive heat capacity with an arbitrary temperature-dependence \cite{ajp2022}. Here,
the body is first brought to an (intermediate) equilibrium state at
some common
temperature which is then connected with the final equilibrium
state via a {\it single} reversible path.
For the sake of completeness, we first describe
this approach in a situation where
the body may be divisible into discrete elements,
or alternately, we have a collection of $n$ bodies,
each in internal equilibrium (Section II).
This approach is extended 
to the continuum case in which the
elements are suitable divisions of a nonuniformly heated
one-dimensional rod, such that each element is in local equilibrium and
the rod has a continuous temperature profile (Section III).
We illustrate the general calculations
using a rod with a constant (temperature-independent) heat capacity in Section IIIA and present an interpolating
model between the discrete and continuum cases in Section IIIB. The case of two bodies where one 
of them has a constant negative heat capacity is described in Section IV with an Appendix treating the general case
of temperature-dependent heat capacity. We conclude in Section V.

\section{Discrete Case}
Following Ref. \cite{ajp2022}, let each element be labelled
by its initial temperature $T_{i}$, while
the final temperature $T_F$ is evaluated from
the constraint of energy conservation:
\begin{equation}
 \sum_{i} \int_{T_{i}}^{T_F} {C_i dT} = 0,
 \label{consen}
\end{equation}
where the sum extends to all the $n$ bodies or elements.
Clearly, to satisfy the above constraint,
all integrals cannot be of the same sign.
This implies that $T_F$ is bounded by
the minimum and the maximum values
 of the given initial temperatures $\{ T_{i}|i=1,...,n \}$:
$T_{\rm min} < T_F <  T_{\rm max}$.

Now, Clausius' result as applied to an individual element
yields the change in entropy as:
$\Delta S_i = \int_{T_{i}}^{T_F} C_i dT /T$,
where $C_i(T) >0$ is the heat capacity
of the element $i$, with some temperature
dependence.
The total entropy change in the body is given by:
\begin{equation}
 \Delta S = \sum_{i} \Delta S_i =  \sum_{i} \int_{T_{i}}^{T_F} \frac{C_i dT}{T}.
\label{dst}
 \end{equation}
In accordance with the Second law, we
expect  $\Delta S > 0$.
However, $\Delta S_i$ is positive
for some values of $i$ (with $T_i< T_F$),
while being negative for others.
Even for the case of two subsystems, the Second law
inequality is not apparent with a generic
heat capacity. Due to $T_2 < T_F < T_1$, one integral
in the sum above is positive while the other one is negative.
The sum may be evaluated for bodies with a specific form of
$C_i(T)$---often taken to be temperature-independent,
and then making use of certain algebraic inequalities \cite{Pyun1974, Lima2015, Mungen2015, Anacleto2016}.
However, a
general argument in this context is usually lacking
(see also \cite{Leff1977}).

It has been shown recently \cite{ajp2022} that the
above total change in entropy is expressible in the form
\begin{equation}
 \Delta_{}^{} {S} =
 \sum_{i}^{} \int_{T_{f}^{}}^{T_{F}^{}}  \frac{C_i dT}{T},
 \label{dsumfF}
 \end{equation}
where $T_f$ defines a new temperature of the whole body,
satisfying $T_{\rm min} <T_f < T_F < T_{\rm max}$.
In this form, each integral
in the above sum is positive-valued, making the Second law
inequality a self-evident result.
To demonstrate Eq. (\ref{dsumfF}),
consider an alternate two-step process
and rewrite Eq. (\ref{dst}) as follows.
\begin{equation}
 \Delta_{}^{} {S} =
 \sum_{i}^{}\int_{T_{i}^{}}^{T_{f}^{}} \frac{C_i dT}{T}
+ \sum_{i}^{} \int_{T_{f}^{}}^{T_{F}^{}}  \frac{C_i dT}{T}.
\label{dsFn}
\end{equation}
In the above, each sum signifies a distinct step of
the alternate process. The first sum denotes
 a reversible
process in which the total entropy
of the body does not change while
different elements are brought to a common
temperature $T_f$. This may be done by
running infinitesimal reversible cycles
between elements, thus gradually reducing the
temperature difference between them while
extracting maximum work in each cycle.
So, the first sum above vanishes by definition.
\begin{equation}
 \sum_{i}^{}\int_{T_{i}^{}}^{T_{f}^{}} \frac{C_i dT}{T}
 =0.
 \label{dszero}
\end{equation}
Note, however, that Eq. (\ref{dszero}) is needed to determine $T_f$.
Also, as will become clear later, the final temperature
after the reversible step satisfies $T_f < T_F$.
Therefore, the entropy change must be solely due to the second sum in
Eq. (\ref{dsFn}), thus yielding  Eq. (\ref{dsumfF}).
The work extracted in the first step is
given by
\begin{equation}
 W = \sum_{i}^{}\int_{T_{i}^{}}^{T_{f}^{}} {C_i} dT < 0.
 \label{wout}
\end{equation}
The work is extracted or the energy of
the system decreases in the process, because the final
equilibrium state is at a lower
energy than the initial non-equilibrium state, for a
fixed entropy during the process. This is just the well-known
energy minimization principle defining
the equilibrium state of a system for
a given value of its entropy \cite{Callen}.

The second step is the reversible heating of the body in
which its temperature changes from $T_f$ to $T_F$.
The magnitude of heat input ($Q$) in this step is equal to the work output ($W$)
in the first step. This may be seen by rewriting
Eq. (\ref{consen}) as:
\begin{equation}
 \sum_{i} \int_{T_{i}}^{T_f} {C_i dT} +
 \sum_{i} \int_{T_{f}}^{T_F} {C_i dT}
 = 0,
 \label{consen2}
\end{equation}
which yields,
$Q = \sum_{i} \int_{T_{f}}^{T_F} {C_i dT} = - \sum_{i} \int_{T_{i}}^{T_f} {C_i dT}
=-W > 0$. This is clearly consistent with
the Second law stipulating that the
energy stored in a single work reservoir
may be completely converted into an equal
amount of heat.

The alternate process described above
is compared with the original
isoenergetic process in Fig. 1.
The initial non-equilibrium state 
is represented by point $A$
where the
energy sum is $\sum_i U_i \equiv U_0$ and entropy sum $\sum_i S_i \equiv S_0$, both energy and entropy being measured
relative to some reference values.
In the process ($A\to C$), the total energy
is conserved while the total entropy increases
 to $S_F$. The point
$C$ represents the equilibrium state
for a given energy $U_0$ and temperature $T_F$.
The alternate two-step process takes place in the manner:
$A \to B \to C$. In the first step ($A\to B$),
the total entropy of the body remains fixed, but the energy
is lowered to $U_f$, due to extraction of work
($W=U_f - U_0 <0$).
The body thus arrives at an equilibrium state ($B$) with temperature $T_f$.
Then, the body is reversibly heated ($B \to C$) raising
its temperature from $T_f$ to $T_F$. Note that
during this process, the whole body remains in internal equilibrium
and thus follows the $U(S)$ curve.

As a simple example,
consider bodies with constant, temperature-independent
heat capacities $C_i$.
Then, from the original process and using
Eq. (\ref{consen}), we can calculate $T_F = \sum_i{a_i T_i}$,
where $a_i = C_i / \sum_i C_i$.
In the alternate process, using the entropy conserving condition
(Eq. (\ref{dszero})), we obtain:
$T_f = \exp \left( \sum_i a_i \ln T_i \right)$.
The change in entropy, from Eq. (\ref{dsumfF}), is given by:
$\Delta S  = S(T_F) - S(T_f)
 = \sum_i \left[ S_i (T_F) - S_i (T_f) \right]$,
  which can be put in the form:
  $\Delta S   = (\sum_i C_i) \ln ({T_F}/{T_f})$.
  Since, the logarithm is a concave function,
  we may apply the Jensen's inequality
  \cite{Jensen,Plastino1997} to verify
  $T_F > T_f$, and hence $\Delta S > 0$.

We close this section with a few general remarks.
The second step consisting of adding heat equal in
magnitude to the extracted work, 
may be accomplished in two different ways. 
Either we supply 
work irreversibly, thus turning
it into an equivalent amount of heat, or
we can add heat {\it reversibly}, by bringing the
body in contact with a series of heat reservoirs
with temperatures ranging from $T_f$ to $T_F$.
Although, the final state obtained for the body
is the same in both cases, there is a subtle difference
of the physics.
If we add work dissipatively to the
body, the entropy of the body is said to increase due to
entropy production. On the other hand, if we
add heat reversibly, then the entropy of the
body increases since heat is being directly
added to it.
Finally, the alternate process highlighted
in Fig. 1  does
not increase the entropy of the universe,
although there is an increase in the entropy
of the body. Thus, even as the body 
achieves the same final state, the
process is reversible, in contrast
to the original isoenergetic process.
In a sense, this is expected since
any irreversible process can be made
reversible provided that sufficient resources
(in the form of reversible heat engines,
series of heat reservoirs and so on) are
made available.

\begin{figure}[ht]
 \includegraphics[width=8cm]{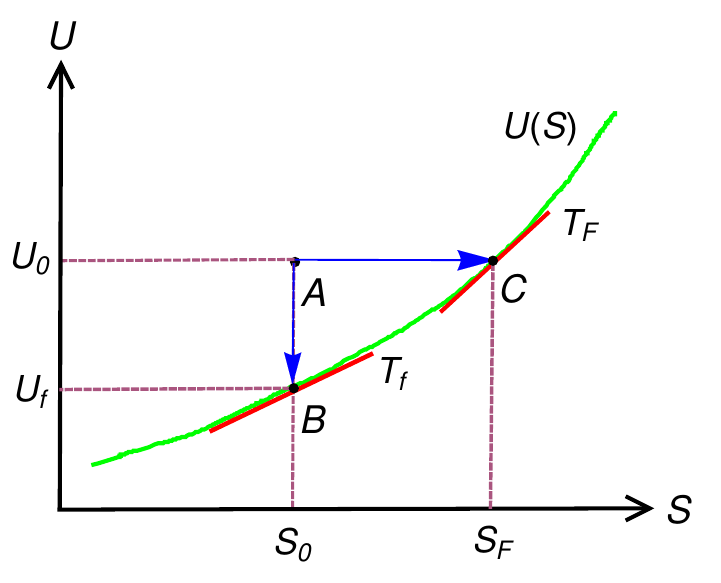}
 \caption{The (green) curve $U(S)$ represents the
 fundamental relation of the body at
 equilibrium, at a given
 volume $V$ and the number of moles $N$.
 Point $A$ which lies {\it off} this curve,
 represents the initial non-equilibrium state.
 We can go from state $A$ to the equilibrium
 state $C$ by preserving the initial energy,
 but increasing the entropy of the system --- in
 accordance with the Second law. Alternately,
 we can go from state $A$ to another equilibrium state
 $B$ (lying on the equilibrium curve) by keeping
 the entropy fixed, but lowering the energy
 by extracting work. The (red) tangents at
 points $B$ and $C$ denote the final
 equilibrium temperatures $T_f$ and $T_F$, respectively.
 One of the conditions of stable equilibrium is
 $(\partial^2 U/ \partial S^2)_{V,N}^{} \geq 0$,
which is equivalent to the positivity of heat capacity
 at constant volume and ensures that $T_f < T_F$.}
 \label{fig1}
\end{figure}

\section{Continuous case: Thermalization of a rod}
We now consider a heat conducting rod of length $L$ whose one end ($x=0$) is in contact with a heat reservoir at temperature $T_1$. The other end ($x=L$) is similarly maintained at temperature $T_2 \; (< T_1)$. Heat flows from the hot to the cold end. A steady state is achieved in which the
temperature profile along the length of the rod is given by a
monotonic increasing function
$T_{\rm in}(x)$, satisfying $T_2 \leq T_{\rm in}(x) \leq T_1$,
where $0\leq  x \leq L$.
The rod may be regarded as a continuum of elements, each of length $dx$.
Accordingly, the sum over index $i$ of the elements,
performed in the previous section,
may be replaced by an integration over the position variable $x$.
The rod, with the initial profile
$T_{\rm in}(x)$,  is now  thermally isolated from the surroundings.
This results in heat flow between the elements due to
their unequal temperatures.
After a sufficiently long
time, a uniform
temperature ($T_F$) is obtained along the length
of the rod and the flow of heat stops.

Let $C(x, T) >0$ be the heat capacity  per unit length at position $x$ of the rod with local temperature $T(x)$. We define a real-valued function of the common temperature $T$ of the rod, given by:
\begin{equation}
\mathscr{Q}(T) = \int_{0}^{L}\!dx
 \int_{T_{\rm in} (x)}^{T}C(x,z)\,dz.
 \end{equation}
 Physically, it denotes the net heat exchanged by the rod in a process with the final temperature $T$. Since ${d\mathscr{Q}}/{dT} = \int_{0}^{L}C(x,T)dx \;> 0$,  it implies \(\mathscr{Q}\) is an increasing function of \(T\).
Thus, the function admits only one zero,
i.e. $\mathscr{Q}(T_{F}) = 0$, where \(T_{F}\) is a generalized mean temperature,  satisfying $T_2 \leq T_F \leq T_1$.
The condition
\begin{equation}
\mathscr{Q}(T_F) = \int_{0}^{L}\!dx
 \int_{T_{\rm in} (x)}^{T_F}C(x,z)\,dz =0,
 \label{constrod}
 \end{equation}
also implies conservation of energy of the rod
in an adiabatic process when the work involved is also zero.

Let \(\Delta \mathscr{S}(T)\) be another real-valued function, defined as:
\begin{equation}
    \Delta \mathscr{S}(T) = \int_{0}^{L}\!dx
 \int_{T_{\rm in} (x)}^{T} \frac{C(x,z)}{z}\,dz.
 \label{sT}
\end{equation}
Since the temperature is defined to be positive, $\Delta \mathscr{S}(T)$ is also an increasing function of $T$.
The function \(\Delta \mathscr{S}(T)\) evaluates the entropy change of the total system (rod) from the initial non-equilibrium state to the final equilibrium state at a common temperature \(T\).
Thus, \(\Delta \mathscr{S}(T_F)\) defines the change in entropy
of the rod undergoing isoenergetic thermalization.

Now, since $T_{\rm in} (x)$ is a continuous, monotonic function of $x$,
we can define a position $x=r$ with
\( T_{\rm in}(r) = T_{F}\) such that the elements in the $x$-range $[0,r]$ have their initial temperatures {\it lower} than $T_F$, while the elements in the range $(r,L]$ are initially
at temperatures {\it higher} than $T_F$. Note that the value $r$ itself depends on the heat capacity function and initial temperature profile.   Then, we can write
\begin{equation}
    \Delta \mathscr{S}(T_{F}) = \int_{0}^{r}\!dx
 \int_{T_{\rm in} (x)}^{T_F} \frac{C(x,z)}{z}\,dz + \int_{r}^{L}\!dx
 \int_{T_{\rm in} (x)}^{T_F} \frac{C(x,z)}{z}\,dz
 \label{stf}
\end{equation}
Thus, the first term above is positive while the second one is negative.
Similarly, it is convenient to write Eq. (\ref{constrod}) in the following form:
\begin{equation}
\mathscr{Q}(T_F) = \int_{0}^{r}\! dx
 \int_{T_{\rm in} (x)}^{T_F}C(x,z)\,dz +
 \int_{r}^{L}\!dx
 \int_{T_{\rm in} (x)}^{T_F}C(x,z)\,dz =0.
 \label{qtf2}
 \end{equation}
Now, invoking the decreasing nature of the function $1/z$, we obtain
\begin{align}
    \int_{0}^{r} \!dx
 \int_{T_{\rm in}(x)}^{T_F}\frac{C(x,z)}{z}dz &>
       \frac{1}{T_F}\int_{0}^{r}\!dx \int_{T_{\rm in}(x)}^{T_F}C(x,z) dz \nonumber \\
     & =-\frac{1}{T_F}\int_{r}^{L}\!dx \int_{T_{\rm in}(x)}^{T_F}C(x,z) dz \nonumber \\
    & = \frac{1}{T_F}\int_{r}^{L}\!dx \int_{T_F}^{T_{\rm in}(x)}C(x,z) dz
    > \int_{r}^{L}\!dx \int_{T_F}^{T_{\rm in}(x)}\frac{C(x,z)}{z} dz,
\end{align}
where we used Eq. (\ref{qtf2}) in
the second step above. Thus, we can write Eq. (\ref{stf})
as
\begin{equation}
   \Delta \mathscr{S}(T_F) = \int_{0}^{r}\!dx \int_{T_{\rm in}(x)}^{T_F}\frac{C(x,z)}{z} dz
    -\int_{r}^{L}\!dx \int_{T_F}^{T_{\rm in}(x)}\frac{C(x,z)}{z} dz >0,
\end{equation}
or
\begin{equation}
    \Delta \mathscr{S}(T_F) = \int_{0}^{L}\!dx \int_{T_{\rm in}(x)}^{T_F}\frac{C(x,z)}{z} dz > 0.
\end{equation}
The above analysis proves that the function \( \Delta \mathscr{S}\) evaluated at \(T_{F}\) is strictly positive.
Since \(\Delta \mathscr{S}(T)\) in Eq. (\ref{sT}) is an increasing function, there must exist a unique {\it lower} value of the temperature \(T_f (< T_{F})\) at which the function \(\Delta \mathscr{S}(T)\) admits a zero \cite{Cashwell67, Landsberg1980g}:
\begin{equation}
    \Delta \mathscr{S}(T_f) = \int_{0}^{L}\!dx \int_{T_{\rm in}(x)}^{T_f}\frac{C(x,z)}{z} dz =0.
    \label{sttz}
\end{equation}
Upon writing $\Delta \mathscr{S}(T_F)$ in the following form
\begin{equation}
    \Delta \mathscr{S}(T_F) = \int_{0}^{L}\!dx \int_{T_{\rm in}(x)}^{T_f}\frac{C(x,z)}{z} dz +
 \int_{0}^{L}\!dx \int_{T_f}^{T_F}\frac{C(x,z)}{z} dz >0,
 \label{sttf}
\end{equation}
we obtain from Eqs. (\ref{sttz}) and (\ref{sttf}):
\begin{equation}
    \Delta \mathscr{S}(T_F) = \int_{0}^{L}\!dx \int_{T_f}^{T_F}\frac{C(x,z)}{z} dz > 0.
\end{equation}
We recall that \(T_F\) is evaluated from the energy-conserving
condition (\ref{constrod}), while \(T_f\) follows from
the entropy-conserving condition (\ref{sttz}). The change in
internal energy in the first step of the alternate process
is extracted in the form of work:
\begin{equation}
 \mathscr{W} = \int_{0}^{L}\!dx
 \int_{T_{\rm in} (x)}^{T_f} {C(x,z)}\,dz <0.
\end{equation}
The amount of heat added in the second step is given by:
\begin{equation}
\mathscr{Q}' = \int_{0}^{L}\!dx
 \int_{T_{f}}^{T_F}C(x,z)\,dz >0.
 \label{rodheat}
 \end{equation}
which by virtue of the condition (\ref{constrod}) implies $\mathscr{Q}' = -\mathscr{W}$.

\subsection{A special case}
After demonstrating the Second law inequality in a rod
with a general heat capacity function $C(x,T)$ and a
monotonic initial profile $T_{\rm in}(x)$,
we illustrate our results for a special case. Consider
the rod with a constant heat capacity per unit length i.e. \(C(x,T) = C\)
\cite{comment_cx}.
First, treating thermalization as an adiabatic
process, the constraint of energy conservation (Eq. (\ref{constrod}))
yields:
\begin{equation}
 T_F = \frac{1}{L} \int_{0}^{L} {T_{\rm in}(x)} dx
 \equiv \langle T_{\rm in}(x) \rangle,
 \label{TFgen}
\end{equation}
where $\langle \cdot\cdot \rangle$ denotes the mean value over $x$.
The change in the entropy of the rod is evaluated as:
 \begin{align}
   \Delta \mathscr{S} & = C  L \ln T_F - C \int_{0}^{L} \ln T_{\rm in}(x) \;   dx.
   \label{dSln0}
  \end{align}
Next, we consider the alternate two-step process.
First, a reversible process is done on the rod such that the
rod attains the common temperature $T_f$.
For the rod with a constant heat capacity per unit length ($C$),
the entropy conserving condition (Eq. (\ref{sttz})) yields:
\begin{equation}
 T_f = \exp  \langle \ln T_{\rm in}(x)    \rangle.
 \label{Tf}
\end{equation}
Using the above expression in Eq. (\ref{dSln0}), we can write:
\begin{equation}
\Delta \mathscr{S} =  C L\ln\frac{T_F}{T_f} > 0,
\label{delsfF}
\end{equation}
since $T_F > T_f$, due to the application of Jensen's inequality:
  $\ln \langle T_{\rm in}(x) \rangle  > \langle \ln T_{\rm in}(x)\rangle$.
The amount of work is given by $\mathscr{W} = C L(T_f- T_F) <0 $
which is equal in magnitude to the heat added in the second step.

Note that we have not yet assumed
any specific form for the initial temperature profile.
Taking
a linear temperature profile \cite{lineartx}, given by
$T_{\rm in} (x) = T_1 - (T_1-T_2)x/L$, we find
\begin{equation}
T_F = \frac{T_2+T_1}{2}, \quad
T_f = \frac{1}{e}\left(\frac{T_1^{T_1}}{T_2^{T_2}}\right)^\frac{1}{T_1 - T_2}.
\end{equation}
The first formula gives the arithmetic mean while the second one is known
as the identric mean \cite{Stolarsky1975,Sandor1990}  of two numbers $T_1 \neq T_2 >0$.
The former is greater than the latter ($T_F > T_f$), with equality only
if $T_1=T_2$.

Finally, from Eq. (\ref{delsfF}),  the entropy change may be explicitly
written in the form:
 \begin{equation}
 \Delta \mathscr{S} = C L
 \left( 1+ \ln T_F - \frac{T_1 \ln T_1 - T_2 \ln T_2}{T_1 - T_2}\right) > 0,
 \label{dSF}
 \end{equation}
which may also  be derived by applying the Clausius' theorem
to the isoenergetic process \cite{Zemanskybook}.

\subsection{An interpolating model}
It can be expected that by refining 
a discrete partition of elements, one may approach the 
continuum model. In the following, we study 
such an interpolating model for the case of the 
rod with a constant heat capacity. 
We show that $T_f$, given by the geometric 
mean of the temperatures 
of $n$ elements,  approaches the identric mean of 
their highest and the lowest temperatures 
as $n\to \infty$. 

Let the rod be composed of $n$ elements arranged in a
row from $x=0$ to $x=L$.
The temperature of the extremal
elements are $T_1$ and $T_n$, being the highest and
the lowest temperatures in the given initial 
profile $\{ T_i \}$. 
We assume a discrete profile for the temperatures, 
such that the temperature of the $k$-th intermediate
 element is given by:
\begin{equation}
 T_k = T_1 - \frac{T_1-T_n}{2n}(2k-1), 
\end{equation}
where $k = 2,3,...,n-1$, with  $n \geq 3$. 
For example, with $n=3$, $T_2 = (T_1 + T_3)/2$. 
With $n=4$, $T_2 = (5T_1 + 3T_4)/8$
and $T_3 = (3T_1 + 5T_4)/8$, thus satisfying
$T_1 > T_2 > T_3 > T_4$ (see Fig. 2).

As we have assumed a homogeneous rod with a constant 
heat capacity $C$, the heat capacity of each 
element is equal to $C/n$. 
Now, the entropy-conserving temperature, $T_{f}^{(n)}$, 
resulting from the maximum work extraction
from the $n$-elements rod, is given by 
the geometric mean of all the $T_i$s. 
Thus, 
\begin{equation}
 T_{f}^{(n)} = \left(T_1\cdot \prod_{k=2}^{n-1} T_k\cdot T_n\right)^{1/n},
\end{equation}
which may be rewritten in the form:
\begin{equation}
 T_{f}^{(n)} = \exp\left[ \frac{1}{n} \left(\log T_1+  \sum_{k=2}^{n-1} \ln T_k + \ln T_n \right)\right],
\end{equation}
where the exponent is the arithmetic mean of $\ln T_i$ with 
$i=1,2,...n$. In the limit, $n\to \infty$, we expect to approach 
the continuous profile, and so the arithmetic mean 
may be replaced with an integral mean of $\ln T$ over a uniform
measure, $1/(T_1-T_n)$. Therefore, 
\begin{equation}
 T_{f}^{(\infty)} = \exp\left[\frac{1}{T_1-T_n} \int_{T_n}^{T_1}\! \ln T \; dT \right].
 \label{tidmean}
\end{equation}
Upon evaluating the integral and simplifying, we find that
$T_{f}^{(\infty)}$ is the identric mean of 
the highest and the lowest temperatures that we found earlier. The inset of Fig. 2 shows $T_{f}^{(n)}$ approaching the identric mean for large $n$ values. 
Now, the initial 
entropy of the $n$-elements rod  is 
given by $S_{\rm in} = C \ln T_{f}^{(n)}$ with respectively to some reference value, 
implying that the initial entropy of the 
rod increases as a function of $n$ and so obtains its 
maximum for the 
continuous rod. On the other hand, the 
energy-conserving temperature $T_F =(T_1 + T_n)/2$
does not vary with the number of partitions $n$,
implying that the final 
entropy of the rod is given by $S_{\rm fin} = 
C \ln T_{F}^{}$, independent of the index $n$.
We thus observe that the amount of 
entropy generated during isoenergetic thermalization
by an $n$-elements rod 
decreases as $n$ increases. 

\begin{figure}
 \includegraphics[width=8cm]{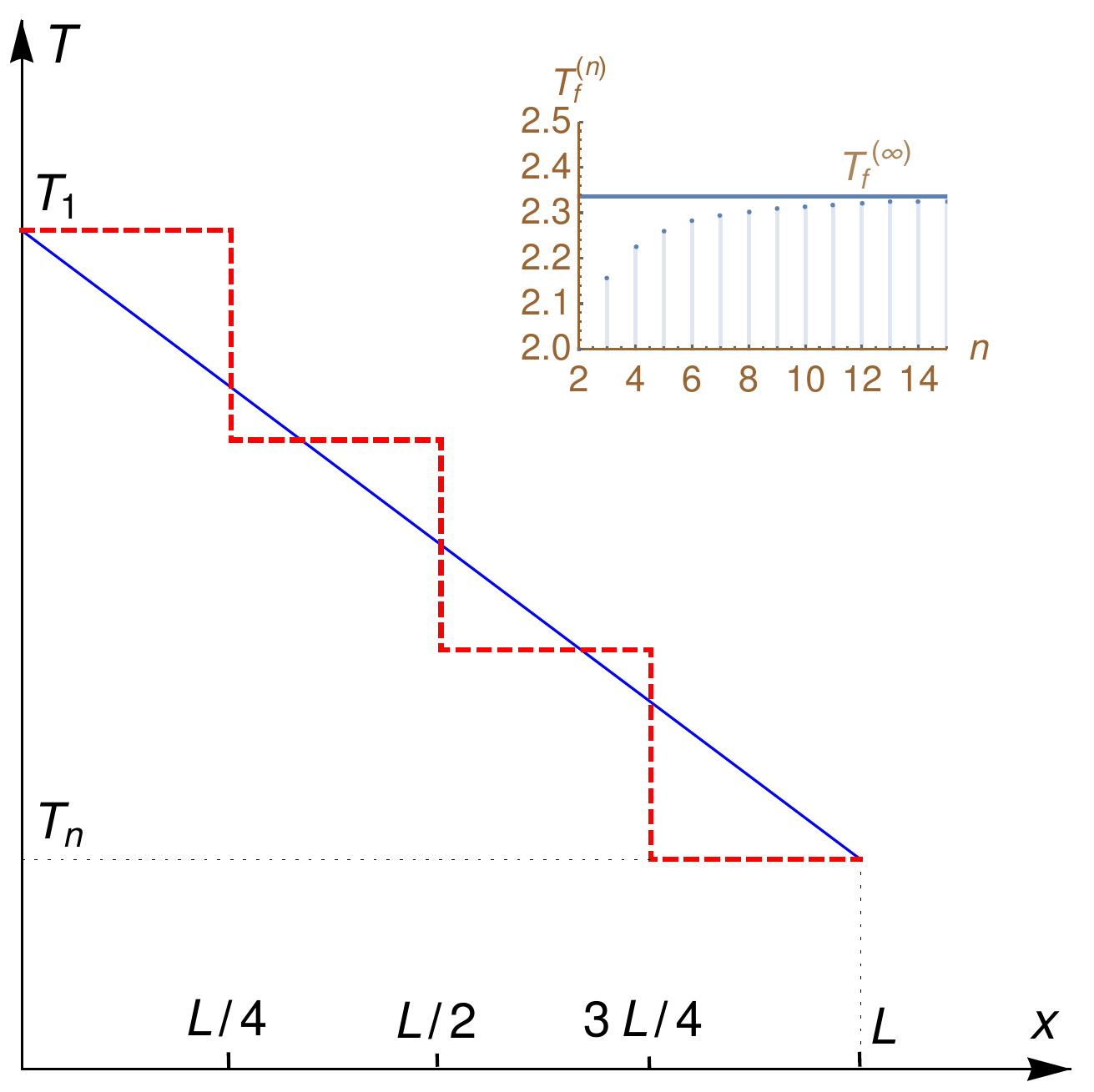}
 \caption{Discretization of a linear temperature
 profile (blue sloping line) of the homogeneous
 rod of length $L$,
 for given extremal temperatures 
 $T_1$ and $T_n$. The inset shows how the 
 entropy-preserving temperature $T_{f}^{(n)}$ approaches the identric
 mean temperature $T_{f}^{(\infty)}$,  Eq. (\ref{tidmean}), 
 as the number of elements becomes large.}
\end{figure}

\section{The case of negative heat capacity}
In the above, we have assumed positive heat capacities for all
the elements composing the nonequilibrium system.
Although a positive heat capacity is required for a system
to be in stable thermal equilibrium with a heat reservoir,
negative heat capacities have been reported
in literature \cite{Lyndenbell, SchmidtNa, Borderie}. An
isolated system or the one described within microcanonical
ensemble can exhibit a negative heat capacity.
Long-range forces also lead to such a
feature \cite{Ruffo}.
 This motivates
the analysis of phase transitions in such systems and
the existence of mutual equilibrium between systems having heat
capacities with different signs. For simplicity, we consider here
the case of constant heat capacities. 
The general case of temperature-dependent heat capacities
is discussed in the Appendix.
Thus, two systems  at temperatures $T_1 > T_2$
and heat capacities satisfying $C_2 < 0 < C_1$ where 
$C_1 + C_2 < 0$, can approach mutual equilibrium when
put in mutual thermal contact \cite{Landsberg1987}. This implies that
the process leads to a rise in the total entropy of the two bodies.
On the other hand, if we have $C_2 < 0 < C_1$ and
$C_1 + C_2 > 0$, then an adiabatic thermalization process  leads
to a decrease in the total entropy. Note that a common
temperature $T_F$ satisfying energy conservation
is defined by the equation: 
 $(C_1 + C_2) T_F = C_1 T_1 + C_2 T_2$,
which can be rearranged as
$ T_1 = a T_2 + (1-a) T_F$,
with the weight $a = |C_2|/C_1$ satisfying $0< a < 1$.
Therefore, $T_1$ is a weighted arithmetic mean of $T_2$ and $T_F$ implying that $T_F > T_1 > T_2$.
Thus, in contrast to the case of positive heat
capacities where $T_F$ lies in between the 
high and low initial temperatures, here the 
relevant $T_F$ is larger than both the temperatures. 
Then, it can be shown that an
energy-conserving process establishing a common
temperature $T_F$ leads to a decrease in the total
entropy, and so this process is forbidden.

This motivates the study of the alternate
two-step process in the present case for its feasibility and hence consistency with the 
Second law. To implement it,
 we first perform an entropy-conserving process---supplying work
reversibly to the system so that heat is pumped from body 2 to 1.
As a result, the temperatures of both bodies show a rise, but
since \(C_1 > |C_2|\), the temperature of body 2 catches up with that of  body 1,
and they  achieve a common temperature \(T_f\),
where \(T_f > T_1 > T_2\). This temperature, as earlier, is obtained
from the entropy-conserving condition which yields
\(T_f = T_{1}^{\alpha} T_{2}^{\beta}\), where
 \(\alpha + \beta = 1\) with
\(\alpha = {C_1}/({C_1 + C_2}) >0 \) and \(\beta = {C_2}/({C_1 + C_2}) <0 \). So the work supplied is calculated as:
\begin{align}
    W &=   U_{\rm fin} - U_{\rm in}, \nonumber\\
    &= (C_1 + C_2) T_f - (C_1 T_1 + C_2 T_2), \nonumber\\
    &= (C_1 + C_2)(T_f - T_F).
\end{align}
In the present case, we have \(T_f > T_F \) as the 
arithmetic-geometric mean inequality is reversed. 
So, it is verified that \(W > 0\), or the
work needs to be done on the system.

In the second step, we
put this composite system---having temperature \(T_f\) and a positive
heat capacity ($C_1 + C_2$), in contact with a series of heat reservoirs till its temperature drops to \(T_F\).
The total heat extracted from the system is given by
$Q = (C_1 + C_2) (T_F - T_f) < 0$,
which is equal in magnitude to the work done on the system 
in the first step.  
The system now acheives its initial energy since \(T_F\) is the energy-preserving common temperature. As discussed above, the change in 
the entropy of the two-body system is negative.
However, the reversible heat exchanges with the series
of reservoirs just manage to compensate for this 
decrease of entropy, thus making the whole alternate
 process as reversible. This also makes this 
 process an allowed one 
unlike the originally perceived adiabatic process. 
Note that, in the second step, we may choose
to place the system directly in thermal contact with 
a specific reservoir at temperature $T_F$.
This step would then be irreversible and 
there will be a net rise of the entropy
of the universe in the whole process. 

Summarising this section, 
unlike the previous cases of positive heat capacities, an 
adiabatic thermalization process may not be allowed
between two systems having constant  heat capacities with
opposite signs. 
We have shown that an energy-preserving
common temperature can still be achieved using an alternate process 
(consistent with the Second law) and so
it becomes a
natural choice to achieve mutual equilibrium 
in such cases. Similarly, we
can design an alternate process for the case $C_1, C_2 < 0$ and
$C_1 + C_2 < 0$ that does not allow an adiabatic thermalization.

\section{Conclusions}
Isoenergetic or adiabatic thermalization, from 
an initially nonequilibrium state
to a final equilibrium state yielding  a uniform temperature ($T_F$), 
is a paradigmatic process of classical thermodynamics.
Assuming positive heat capacities of the bodies,
 the Second law is usually demonstrated for
this process  by making use of
algebraic inequalities to infer that 
$\Delta S = S(T_F) - S_{\rm in} >0$. In this paper, 
isoenergetic thermalization is executed as a 
two-step process that 
evokes another fundamental process of thermodynamics,
the so-called maximum work theorem.
The alternate process comprises of reversible work extraction in the first step  
by which the composite body attains a 
common temperature $T_f (< T_F)$,
followed
by the second step of adding heat to the 
body---equal
in magnitude to the extracted work, thus restoring
the body to its initial energy with the same final
temperature $T_F$ as in the isoenergetic process.
The first step yields 
the condition $S_{\rm in} = S(T_f)$. Further, the 
positivity of heat capacity implies that entropy 
is a monotonic increasing function of the temperature
of the body. Thereby,
$T_F > T_f$ inequality directly leads to
the result: $\Delta S = S(T_F) - S(T_f) >0$.
 
The study of the above alternate 
process is extended to a continuous system  in the form of a 
conducting rod prepared in a nonequilibrium state by 
placing its two ends in contact with 
hot and cold reservoirs respectively. The results 
have been illustrated with a homogeneous rod of 
constant heat capacity. Further, 
an interpolating model is analyzed which 
leads from a discrete to a continuous 
temperature profile as the number of 
elements becomes very large, showing that 
the continuous rod generates less entropy
during thermalization than the discrete model. 
 
Finally, it is observed that when one or 
both the bodies have a negative heat capacity, 
an adiabatic thermalization process may lead
to violation of the Second law. It is shown that 
 the alternate process  brings  
 the two bodies to an energy-preserving common temperature, in consistency
 with the Second law. Thus, the 
 alternate process not only provides a useful aid
 for understanding entropy production in adiabatic
 processes with positive heat capacities,
 but also provides feasible  
 processes consistent with the Second law, when 
 the heat capacities are negative. 
Apart from a classical analysis of such basic processes, 
their study in the realm of quantum thermodynamics can provide an 
interesting direction for future investigations.

 \section*{Acknowledgement}
 VN acknowledges the financial support from IISER Mohali during the Int-PhD program.

\section{Appendix: Temperature-dependent negative
heat capacity}
Consider two bodies having temperature-dependent heat capacities $C_1(T) >  0$, $C_2(T) <  0$ such that $C_1(T) + C_2(T) > 0$,
for all temperatures $T>0$.
Let the initial temperatures of the bodies be $T_1$ and $T_2$. Without loss of generality, we assume $T_1 > T_2$.
For $T>  T_1$, the change in total energy is given by
\begin{equation}
 \Delta U(T) = \int_{T_1}^{T} C_1 (z) dz +
  \int_{T_2}^{T} C_2 (z) dz,
  \label{utz}
\end{equation}
where the energy-conserving common temperature
is given by the condition $\Delta U(T_F) =0$ which can
be rewritten as:
\begin{equation}
 \int_{T_2}^{T_1} C_1 (z) dz  =
  \int_{T_2}^{T_F} [C_1 (z) + C_2 (z)] dz.
  \label{utFz}
\end{equation}
From the above relation, we can conclude that
$T_F > T_1 > T_2$ \cite{Landsberg1987}.

Again, for $T> T_1$, the change in the
total entropy of the two-body system is
given by:
\begin{align}
    \Delta S(T) &= \int_{T_1}^{T} \frac{C_1(z)}{z}dz + \int_{T_2}^{T} \frac{C_2(z)}{z}dz \\
    &=  \int_{T_1}^{T} \frac{\left[ C_1(z) + C_2(z) \right]}{z}dz  +  \int_{T_2}^{T_1} \frac{C_2(z)}{z}dz\\
    &\leq  \frac{1}{T_1}   \int_{T_1}^{T} [C_1(z) + C_2(z) ] dz  +  \frac{1}{T_1}\int_{T_2}^{T_1} C_2(z)dz \\
    & \leq  \frac{1}{T_1}  \left(  \int_{T_1}^{T} C_1(z) dz + \int_{T_2}^{T} C_2(z) dz  \right).
\end{align}
In other words, we obtain $\Delta S (T) \leq \Delta U(T) / T_1$.
Therefore, $\Delta S(T_F) \leq 0 $, implying that the process
of adiabatic equilibration is forbidden in this case.
Further, since $\Delta S(T)$ is a monotonic increasing function
due to
\( (d/dT)\Delta S(T) =
[ C_1(T) + C_2(T)]/{T} >0\), therefore the
entropy-preserving temperature $T_f$, defined by \(\Delta S(T_f) = 0\), must be greater than $T_F$. Thus, we have
 the relations:  $T_2 < T_1 < T_F < T_f$.
 Now, we can consider the two-step alternative process
 in which the work is supplied to bring the two
 bodies to a common temperature $T_f$ and then
 an equivalent amount of heat is extracted by
 attaching to a series of reservoirs, thereby
 bringing the common temperature to $T_F$.
 The system comes back to a state with same energy
 ($\Delta U(T_F) =0$), and the total change in
 entropy is zero for the whole
 process, thus allowing the process.

%
%
%


\begin{thebibliography}{7}
\bibitem{comment_cx} In general, we may consider a heterogenous rod with a
temperature independent heat capacity. In that case, $C(x,T) = C(x)$.
This is analogous to the discrete case with different $C_i$ for each body.
\bibitem{lineartx} A linear temperature profile is a
natural consequence of a temperature-independent
thermal conductivity of the rod.
\end{thebibliography}
\end{document}